\documentclass[preprint,showpacs,preprintnumbers,amsmath,amssymb,nofootinbib,superscriptaddress]{revtex4}

\usepackage{graphicx}
\usepackage{dcolumn}
\usepackage{bm}

\begin{document}


\title{Collective flow and two-pion correlations from a relativistic\\
hydrodynamic model with early chemical freeze out
}

\author{Tetsufumi Hirano}
 \email{hirano@nt.phys.s.u-tokyo.ac.jp}
\affiliation{%
Physics Department, University of Tokyo, Tokyo 113-0033, Japan
}%

\author{Keiichi Tsuda}
\affiliation{%
Department of Physics, Waseda University, Tokyo 169-8555, Japan
}%

\date{\today}

\begin{abstract}
We investigate the effect of early chemical freeze-out on radial flow, elliptic flow and HBT radii by using a fully three dimensional hydrodynamic model.
When we take account of the early chemical freeze-out, the space-time evolution of temperature in the hadron phase is considerably different from the conventional model in which chemical equilibrium is always assumed.
As a result, we find that radial and elliptic flows are suppressed and that the lifetime and the spatial size of the fluid are reduced.
We analyze the $p_t$ spectrum, the differential elliptic flow, and the HBT radii at the RHIC energy by using hydrodynamics with chemically non-equilibrium equation of state.
\end{abstract}

\maketitle

\section{Introduction}
\label{sec:intro}

One of the main topics in the physics of relativistic heavy-ion collisions is to give a detailed description of space-time evolution for the hot and dense nuclear matter by using a dynamical model such as kinetic transport models \cite{Bass:2001gb} or hydrodynamics \cite{Huovinen:2002fp}.
Recent experimental data at the RHIC energy suggest that we may see the significant effect of jet quenching in transverse momentum distribution for neutral pions \cite{Adcox:2001jp} or azimuthal asymmetry for charged hadrons \cite{Snellings:2001nf}.
So hydrodynamic simulations of expanding hot and/or dense matter are indispensable in quantitatively estimating the effect of the medium on the jet quenching \cite{Gyulassy:2001kr}.
One of the authors (T.H.) has already built a fully three dimensional hydrodynamic model which describes not only central but also non-central collisions \cite{Hirano:2001eu}.
In contrary to other hydrodynamic models, e.g., (2+1) dimensional models with the Bjorken's scaling solution \cite{Ollitrault:1992bk,Teaney:1999gr,Kolb:2000sd} or (3+1) dimensional models with cylindrical symmetry along the collision axis \cite{Ornik:1996qk,Sollfrank:1997hd,Hung:1998du,Morita:1999vj}, full (3+1) dimensional hydrodynamic models \cite{Amelin:1991cs,Rischke:1995,Nonaka:2000ek,Hirano:2000xa,Osada:2001hw,Magas:2002ge} enables us to obtain the rapidity dependence of particle distribution, elliptic flow, and HBT radii in \textit{non-central} collisions.

Statistical and hydrodynamics-motivated models give us the characteristic temperatures in relativistic heavy-ion collisions.
These temperatures are very useful in understanding what happens in collisions.
From fitting the model calculation of particle ratios for hadrons to the experimental data, the chemical freeze-out temperature $T^{\mathrm{ch}}$ is obtained at each collision energy \cite{Cleymans:1993zc,Braun-Munzinger,Cleymans,Becattini:2000jw,Braun-Munzinger:2001ip,Kaneta:2001ra}.
On the other hand, we can obtain the thermal (kinetic) freeze-out temperature $T^{\mathrm{th}}$ from the slopes of transverse momentum distribution by assuming the radial flow profile \cite{Schnedermann:1993ws,Cleymans,Broniowski:2001we}.
The temperatures obtained above are usually different from each other at the AGS, SPS, and probably RHIC energies: $T^{\mathrm{ch}} \sim 160$-$200$ MeV while $T^{\mathrm{th}} \sim 100$-$140$ MeV. 
The chemical freeze-out parameters $(T^{\mathrm{ch}}, \mu_{\mathrm{B}}^{\mathrm{ch}})$ at various collision energies seem to be aligned on one line in the $T$-$\mu_{\mathrm{B}}$ plane. This line is sometimes called ``chemical freeze-out line" and can be parameterized by the average energy per particle $<E>/<N> \sim 1$ GeV \cite{Cleymans}.
Those analyses indicates that the system undergoes first the chemical freeze-out where the observed particle ratios are fixed and next the thermal freeze-out where the shape of the transverse momentum distribution is fixed \cite{Shuryak:1999zh}.
Since the time scale of the hydrodynamic evolution is comparable with (or smaller than) that of the inelastic collisions between hadrons, the number changing processes are likely to be out of equilibrium.
This is the reason why there are two sequential freeze-out processes in relativistic heavy-ion collisions.
To give a more realistic description of the temporal and spatial behavior of the hot and dense matter, the above pictures should be included in the model. 
This makes us approach the comprehensive understanding of the deconfined matter, the Quark Gluon Plasma (QGP).

In this paper, we incorporate the different freeze-out temperatures, $T^{\mathrm{ch}}$ and $T^{\mathrm{th}}$, into hydrodynamics and discuss how the early chemical freeze-out affects the space-time evolution of fluids and the particle spectra \cite{Hirano:2002qj}.
In the ordinary hydrodynamic calculations, one assumes both chemical and thermal equilibrium and consequently $T^{\mathrm{ch}} = T^{\mathrm{th}}$ which is to be determined from comparison of the slopes of transverse spectrum with the experimental data.
If the system obeys the above picture of early chemical freeze-out, those ordinary hydrodynamic models can hardly reproduce the particle ratios due to the smallness of the chemical freeze-out temperature.
As a result, the number of resonance particles at the thermal freeze-out becomes too small.
The physics at the low transverse momentum is largely affected by resonance decays after thermal freeze-out.
For example, the low $p_t$ enhancement in the transverse momentum spectrum for pions can be explained by the contribution from resonance decays \cite{Ornik:1991dm,Sollfrank:1991xm}.
The slope of the $p_t$ spectrum for pions directly emitted from the freeze-out hypersurface is almost constant.
On the other hand, data for pions at low $p_t$ shows a steeper slope than those in medium $p_t$ region ($\sim$ 1 GeV/$c$).
So one cannot reproduce the low $p_t$ enhancement only from direct pions. Pions from $\rho$, $\omega$, or $\Delta$ decays show steeper $p_t$ spectrum than direct pions.
We naturally explain the low $p_t$ enhancement seen in the experimental data by resonance decays.
Another example is the reduction of the second Fourier coefficient of the azimuthal distribution for pions at low $p_t$ \cite{Hirano:2000eu}. From the exact treatment of the decay kinematics, pions from resonance decays can pretend \textit{out-of-plane} elliptic flow even when the hydrodynamic flow shows \textit{in-plane} elliptic flow.
This dilutes elliptic flow from direct pions.
Therefore the early chemical freeze-out must be included in hydrodynamics in order to analyze not only the particle ratios but also the single particle spectra and the azimuthal asymmetry.

This paper is organized as follows: We construct the equation of state for the hadron phase with and without chemical equilibrium in Sec.~\ref{sec:EOS}.
We parameterize the initial condition for hydrodynamic simulations in the full three dimensional Bjorken coordinate in Sec.~\ref{sec:initial}. 
By using these equations of state and the initial condition, we perform hydrodynamic simulations in full three dimensional space at the RHIC energy and compare space-time evolutions with each other in Sec.~\ref{sec:evolution}.
We analyze the particle spectra and the azimuthal asymmetry at the RHIC energy and discuss the effect of early chemical freeze-out on observables in Sec.~\ref{sec:spectra}. We also analyze the two-pion correlation functions to see the effect on the hydrodynamic evolution in Sec.~\ref{sec:HBT}.
Summary and discussions are given in Sec.~\ref{sec:summary}.

\section{Equations of State}
\label{sec:EOS}

We assume the following picture of space-time evolution for hot and/or dense matter produced in relativistic heavy-ion collisions.
First, the huge number of secondary partons are produced and both chemical and thermal equilibrium among these partons are achieved in the early stage of collisions.
The initial dominant longitudinal flow and the large pressure gradient perpendicular to the collision axis cause the system to expand and cool down.
When the temperature of the system reaches the critical temperature $T_{\text{c}}$, the hadronization starts to occur.
Just after the hadronization finishes, the chemical freeze-out happens at $T^{\text{ch}}(\le T_{\text{c}})$. 
Below $T^{\mathrm{ch}}$, the ratios of the number of observed particle are fixed.
Even after chemical freeze-out, the system keeps thermal equilibrium through elastic scattering.
Finally, all hadrons are thermally frozen at $T^{\text{th}}$($<T^{\text{ch}}$). 
If we neglect dissipation in the space-time evolution of nuclear matter, we can apply the hydrodynamic equations for the perfect fluid, $\partial_\mu T^{\mu \nu}=0$ and $\partial_\mu n_{\text{B}}^\mu=0$, where $T^{\mu \nu} = (E+P)u^\mu u^\nu-Pg^{\mu\nu}$ and $n_{\mathrm{B}}^\mu = n_{\mathrm{B}} u^\mu$ are energy momentum tensor and baryon density current, respectively. $E$, $P$, $n_{\mathrm{B}}$, and $u^\mu$ are energy density, pressure, baryon density, and four fluid velocity.
With the help of thermodynamical identities and the baryon density conservation, the first equation is rewritten in $\partial_\mu s^\mu = 0$, where $s^\mu$ is the entropy current.
These equations mean that a fluid element evolves along an adiabatic path ($n_{\text{B}}/s$ = const.) in the $T$-$\mu_{\text{B}}$ plane.
We assume this fact is approximately valid even below chemical freeze-out line.

When $N$ stable hadrons in the equation of state (EOS) undergo chemical freeze-out across the chemical freeze-out line, we can introduce chemical potentials $\mu_i$ associated with those hadrons.
Then we may construct the EOS in the ($N+2$) dimensional space ($N$ for $\mu_i$ and 2 for $T$ and $\mu_B$).
This causes a serious problem when we numerically simulate a hydrodynamic model with finite resources of memory.
Since the chemical potential of each hadron depends on $T^{\text{ch}}$, $\mu_{\text{B}}^{\text{ch}}$, $T$, and $\mu_{\text{B}}$ during expansion along an adiabatic path, we can restrict ourselves to the EOS in a two dimensional hypersurface $\mu_i=\mu_i (T,\mu_{\mathrm{B}})$ embedded in the $(N+2)$ dimensional space.
Thus we need not prepare such a large dimensional table of the EOS anymore.
When we obtain $\mu_i$ at a point ($T$, $\mu_{\text{B}}$) below the chemical freeze-out line, we need the information at the chemical freeze-out point $(T^{\text{ch}}, \mu_{\text{B}}^{\text{ch}})$ somewhere on the chemical freeze-out line and have to go back to the point along an adiabatic path.
Since the adiabatic path itself is obtained from the thermodynamical variables and $\mu_i(T, \mu_{\text{B}})$, we have to solve the problem self-consistently.
In this paper, we restrict our discussion to the zero baryonic chemical potential where the adiabatic path becomes a trivial one, $n_{\mathrm{B}}/s=0$.
This is a good approximation in Au + Au $\sqrt{s_{NN}} = 130$ GeV collisions in which $\bar{p}/p \sim 0.6$ \cite{pbarpratio,Back:2001qr,Bearden:2001kt,Adcox:2001mf} or the resultant chemical freeze-out parameter $\mu_{\mathrm{B}}^{\mathrm{ch}} \sim$ 50 MeV \cite{Braun-Munzinger:2001ip}.

We construct three models EOS to compare the space-time evolution of fluids. These models describe the first order phase transition between the QGP phase and the hadron phase at $T_{\mathrm{c}} = 170$ MeV.
We suppose the QGP phase is composed of massless u, d, s quarks and gluons and that it is common to three models. The EOS for the QGP phase is $P=(E-4B)/3$, where $B$ is a bag constant specified later.
For the hadron phase, we choose three different models EOS as follows.

\subsection{Chemical Equilibrium}

The first model is an ordinary resonance gas model in which complete chemical equilibrium is always assumed (model CE).
This model is employed for the sake of comparison with the other models.
We include strange and non-strange hadrons up to the mass of $\Delta(1232)$. 
Energy density and pressure are as follows:
\begin{eqnarray}
\label{eq:energy}
E & = &  \sum_i \frac{d_i}{(2 \pi)^3}\int d^3 \bm{p} \frac{\sqrt{\bm{p}^2+m_i^2}}{\exp{[(\sqrt{\bm{p}^2+m_i^2}-\mu_i)/T]}\mp1},\\
\label{eq:pressure}
P & = & \mp \sum_i T \frac{d_i}{(2 \pi)^3}  \int d^3 \bm{p} \log\left\{1\mp\exp\left[-\left(\sqrt{\bm{p}^2+m_i^2}-\mu_i \right)/T\right]\right\},
\end{eqnarray}
where $\mu_i = 0$ due to complete chemical equilibrium in this model.
Here the upper and lower signs correspond to bosons and fermions.
We neglect the excluded volume correction which largely affects the hadronic EOS in the high baryon density region.
From the Gibbs's equilibrium condition at $T_{\mathrm{c}} = 170$ MeV, we obtain the bag constant $B^{1/4}$ = 247.2 MeV and the latent heat $\Delta E \sim $1.7 GeV/fm$^3$.

\subsection{Chemical Freeze Out}

The second model is the simplest one which describes the picture of chemical freeze-out (model CFO). Below $T^{\mathrm{ch}}$, we assume the numbers of \textit{all} hadrons $N_i$ included in the EOS are fixed and that the particle number densities obey $\partial_\mu n_i^\mu = 0$.
We introduce a chemical potential $\mu_i(T)$ associated with each species so that $N_i$ becomes a conserved quantity.
From the conservation of entropy $\partial_\mu s^\mu = 0$, the ratio of the particle number density to the entropy density below the chemical freeze-out temperature obeys
\begin{eqnarray}
\label{eq:RATIO1}
\frac{n_i(T, \mu_i)}{s(T,\{\mu_i\})} = \frac{n_i(T_{\mathrm{ch}},\mu_i=0)}{s(T_{\mathrm{ch}},\{\mu_i\}=0)}
\end{eqnarray}
for all hadrons along the adiabatic path.
Here we assume $T^{\mathrm{ch}} = 170$ MeV, which is consistent with a recent analysis based on a thermal model at the RHIC energy \cite{Braun-Munzinger:2001ip}.
From Eq.~(\ref{eq:RATIO1}), we obtain a chemical potential as a function of temperature for each hadron. 
The $\mu_i(T)$ ensures to keep the ratios of the number of each hadron throughout the space-time evolution of a fluid element without explicitely solving $\partial_\mu n_i^\mu = 0$.
All chemical potentials are functions of temperature, so the thermodynamical variables depend only on temperature even after chemical freeze-out.

\subsection{Partial Chemical Equilibrium}

The third model represents a more realistic EOS than the second one. The following model is first discussed in Ref.~\cite{Bebie:1992ij}. The observed particle numbers are always composed of the contribution from direct particles and resonance decays, i.e., $\bar{N}_\pi = N_\pi + \sum_{i\neq \pi} \tilde{d}_{i \rightarrow \pi} N_i$.
Here $\tilde{d}$ is an effective degree of freedom which is a product of the degeneracy $d$ and the branching ratio $B$.
So some elastic processes with large cross sections (e.g., $\pi\pi \rightarrow \rho \rightarrow \pi\pi$, $\pi N \rightarrow \Delta \rightarrow \pi N$, $\pi K \rightarrow K^* \rightarrow \pi K$) can be equilibrated even below $T^{\mathrm{ch}}$ \cite{Rapp} as long as the equality 
\begin{eqnarray}
\frac{\bar{n}_i(T,\mu_i)}{s(T,\{\mu_i\})} = \frac{\bar{n}_i(T_{\mathrm{ch}},\mu_i=0)}{s(T_{\mathrm{ch}},\{\mu_i\}=0)}
\end{eqnarray}
is kept instead of Eq.~(\ref{eq:RATIO1}).
We regard $\pi$, $K$, $\eta$, $N$, $\Lambda$ and $\Sigma$ as ``stable" particles\footnote{
Here we use a term ``stable'' when the lifetime of a hadron is much longer than that of the fluid ($\sim 10$ fm/$c$).
}
 and that all chemical potentials can be represented by chemical potentials associated with these stable particles, e.g., $\mu_\rho = 2\mu_\pi$, $\mu_{K^*} = \mu_\pi + \mu_K$, $\mu_\Delta = \mu_\pi + \mu_N$, and so on. Thus the third model describes the \textit{partial} chemical equilibrium (model PCE) even below $T^{\mathrm{ch}}$ \cite{Bebie:1992ij}.
It should be noted that, after chemical freeze-out, $\mu_N = \mu_{\bar{N}} \enskip (\neq 0)$ with keeping baryonic chemical potential $\mu_B = 0$ in our model.

The model PCE employed here is not the only one to describe the partial chemical equilibrium. There may be other choices for stable particles or other processes to be equilibrated in this model. Various models should be checked by future precise experimental data of particle ratios.

\subsection{Chemical Potentials and Equations of State}

\begin{figure}
\includegraphics[width=0.7\textwidth]{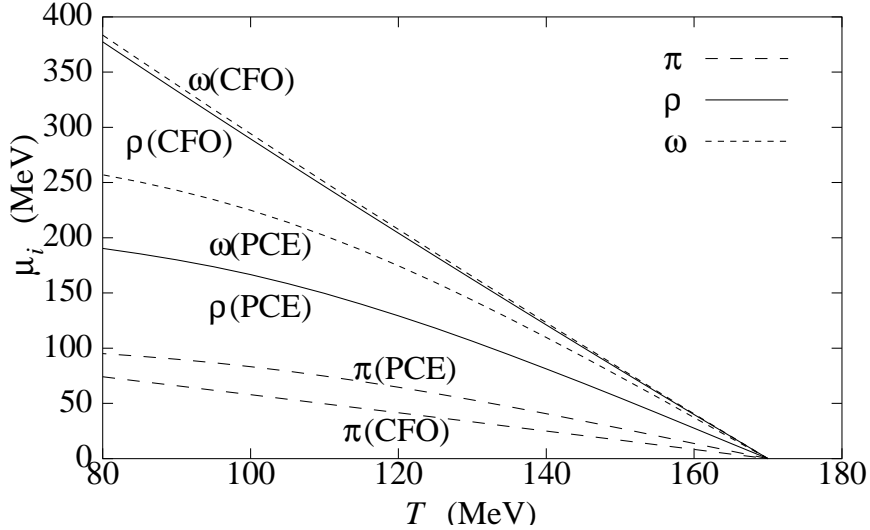}
\caption{\label{Fig1}Chemical potentials for pions (dashed lines), $\rho$ mesons (solid lines), and $\omega$ mesons (dotted lines) for the models CFO and PCE.}
\end{figure}

Figure \ref{Fig1} shows the chemical potentials for $\pi$, $\rho$, and $\omega$ mesons for the models CFO and PCE.
The difference of chemical potentials between hadrons depends only on its mass in the model CFO, so $\mu_\omega(T)$ behaves like $\mu_\rho(T)$ due to the small mass difference.
Both are almost linearly increasing with decreasing temperature. On the other hand, these chemical potentials differ from each other in the model PCE. It results from each elementary process, i.e., $\rho \leftrightarrow \pi\pi$ ($\mu_\rho = 2\mu_\pi$) and $\omega \leftrightarrow \pi\pi\pi$ ($\mu_\omega = 3 \times 0.88 \mu_\pi$), where the branching ratios are from Ref.~\cite{PDG}.

\begin{figure}
\includegraphics[width=0.6\textwidth]{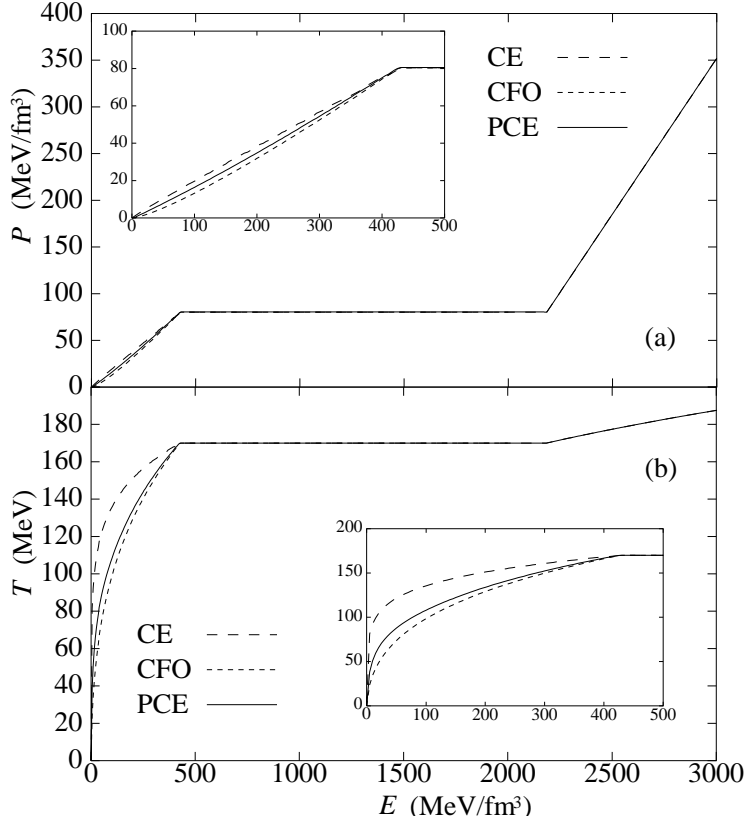}
\caption{\label{Fig2}(a) Pressure as a function of energy density. (b) Temperature as a function of energy density. The dashed, dotted, and solid lines correspond to the models CE, CFO, and PCE.}
\end{figure}

One can easily evaluate EOS's for these models by inserting chemical potentials obtained above into Eqs.~(\ref{eq:energy}) and (\ref{eq:pressure}).
We represent pressure and temperature as functions of energy density for three models in Fig.~\ref{Fig2}. We find in Fig.~\ref{Fig2} (a) that pressure as a function of energy density is similar to each other. On the other hand, temperature as a function of energy density in the models CFO or PCE in Fig.~\ref{Fig2} (b) is deviated from the model CE. Since the resonance population of the models CFO or PCE is larger than that of the model CE due to the chemical freeze-out, the energy density at a fixed temperature in the hadron phase is also large in those models. 
Conversely, temperature in the models CFO or PCE at a fixed energy density is \textit{smaller} than in the model CE. 
This fact is very important in qualitatively understanding the difference of the space-time evolution of fluids among three models as shown in Sec.~\ref{sec:evolution}.

\section{Initial Conditions}
\label{sec:initial}

We numerically solve the hydrodynamic equation in the full three dimensional Bjorken coordinate ($\tau$, $\eta_{\mathrm{s}}$, $x$, and $y$) which is relevant to analyze heavy-ion collisions at the collider energies.
Here $\tau = \sqrt{t^2-z^2}$ and $\eta_s=(1/2)\log[(t+z)/(t-z)]$ are the proper time and the space-time rapidity, respectively. The $x$ axis is parallel to the impact parameter vector and the $y$ axis is perpendicular to the $x$ axis in the transverse plane.
As one of the authors (T.H.) has already pointed out \cite{Hirano:2001eu}, the main reason to employ the Bjorken coordinate rather than the Cartesian coordinate is a practical one: This considerably reduces numerical efforts such as the long lifetime of fluids ($\sim 100$ fm/$c$) and less numerical accuracy near the light cone. In addition to this reason, there is also a physical reason why we avoid to employ the Cartesian coordinate.
When one simulates the space-time evolution in the Cartesian coordinate, one gives the initial condition at a constant $t_0$.
If $t_0$ is regarded as the thermalization time, this initial condition implies that the thermalization occurs first from the forward (backward) space-time rapidity in $\tau$-$\eta_{\mathrm{s}}$ coordinate.\footnote{Supposing $t_0 = 1$ fm/$c$, the corresponding initial \textit{proper} time becomes $\tau_0 = t_0/\gamma_N \sim 0.01$ fm/$c$ near the beam rapidity region at RHIC energies.}
This is somewhat unrealistic because the multiplicity in the forward rapidity region is much smaller than the one at midrapidity.
This may cause a crucial problem when one discusses the rapidity dependence of radial and elliptic flows at the collider energies, since they are sensitive to the thermalization of the system.

We choose initial conditions in the Bjorken coordinate so as to reproduce pseudorapidity distribution in Au+Au 130$A$ GeV central (0-6 \%) collisions obtained by the PHOBOS Collaboration \cite{Back:2001bq}. At the initial time $\tau_0=0.6$ fm/$c$, the initial energy density for central collisions can be factorized as
\begin{eqnarray}
E(x,y,\eta_{\mathrm{s}}) & = & E_{\mathrm{max}} W(x,y;b) H(\eta_{\mathrm{s}}).
\end{eqnarray}
We assume the transverse profile function $W(x,y;b)$ scales with the impact parameter $b$ in proportion to the number of binary collisions \cite{Kolb:2001qz}
\begin{eqnarray}
\label{eq:Tprofile}
W(x,y;b) & \propto & T_{+}T_{-}, \qquad T_{\pm}=T(x \pm b/2,y),
\end{eqnarray}
where $T(x,y)$ is a thickness function with the standard Woods-Saxon parameterization for nuclear density
\begin{eqnarray}
T(x,y) = \int dz \rho(x,y,z), \qquad \rho(x,y,z) = \frac{\rho_0}{\exp[(\sqrt{x^2+y^2+z^2}-R_0)/\delta]+1}.
\end{eqnarray}
Here $\rho_0 = 0.17$ fm$^{-3}$ is the saturation density, $\delta=0.54$ fm is the diffuseness parameter, and $R_0 = 1.12A^{1/3}-0.86A^{-1/3}$ fm is the nuclear radius.
The proportional constant in Eq.~(\ref{eq:Tprofile}) is fixed from the condition $W(0, 0; 0) = 1$.
The longitudinal profile function $H(\eta_{\mathrm{s}})$ is characterized by two parts \cite{Hirano:2001eu,Morita:1999vj,Hirano:2001yi}: it is flat near $\eta_{\mathrm{s}} \sim 0$ and smoothly connects to vacuum as a half part of the Gauss function in the forward and backward space-time rapidity regions 
\begin{eqnarray}
H(\eta_{\mathrm{s}}) & = & \exp\left[-\frac{(\mid \eta_{\mathrm{s}}-\eta_{\mathrm{s0}} \mid - \eta_{\mathrm{flat}}/2)^2}{2\eta_{\mathrm{Gauss}}^2} \theta(\mid \eta_{\mathrm{s}} -\eta_{\mathrm{s0}}\mid - \eta_{\mathrm{flat}}/2)\right].
\end{eqnarray}
The length of a flat region $\eta_{\mathrm{flat}}$ and the width of the Gauss function $\eta_{\mathrm{Gauss}}$ are adjustable parameters to be determined by the experimental data, especially (pseudo)rapidity distribution.
In symmetric collisions with the vanishing impact parameter, we expect the symmetry $E(x,y,-\eta_{\mathrm{s}}) = E(x,y,\eta_{\mathrm{s}})$ holds at the initial time.
On the other hand, in non-central (or asymmetric) collisions, we can shift the energy density by $\eta_{\mathrm{s0}}$ which is identified with the center of rapidity for each transverse coordinate \cite{Sollfrank:1998js}
\begin{eqnarray}
\eta_{\mathrm{s0}}(x,y;b) = \frac{1}{2}\log \left[\frac{(T_{-} +T_{+})\gamma_N + (T_{-}-T_{+})\gamma_N v_N}{(T_{-}+T_{+})\gamma_N - (T_{-}-T_{+})\gamma_N v_N}\right],
\end{eqnarray}
where $v_N$ and $\gamma_N$ are, respectively, velocity and Lorentz $\gamma$ factor of an incident nucleon in the center-of-mass system.
For illustrations of the initial energy density, see Ref.~\cite{Hirano:2001eu}.
The initial longitudinal flow is the Bjorken's solution \cite{Bjorken:1983qr}, i.e., the fluid rapidity $Y_{\mathrm{f}}(\tau_0)$ is equal to the space-time rapidity $\eta_{\mathrm{s}}$. It should be noted that this is merely an initial condition and that $Y_{\mathrm{f}} \neq \eta_{\mathrm{s}}$ after initial time due to the pressure gradient directed to the $\eta_{\mathrm{s}}$ axis. The transverse velocities vanish at $\tau_0$ and are to be generated only by the transverse pressure gradient.

Initial parameters in hydrodynamic simulations are so chosen as follows:
$E_{\mathrm{max}}= 35$ GeV/fm$^3$, $\eta_{\mathrm{flat}} = 5.8$, $\eta_{\mathrm{Gauss}} = 0.2$, and $b=2.4$ fm. 
These values lead us to reproduce the pseudorapidity distribution in Au + Au central collisions at $\sqrt{s_{NN}} = 130$ GeV observed by the PHOBOS collaboration \cite{Back:2001bq}.
The parameters are adjusted for the model PCE with $T^{\mathrm{th}} = 140$ MeV. We also use the same values for the other models EOS for the sake of comparison, although this causes the pseudorapidity distribution slightly deviated from the experimental data.

\section{Time Evolution of Fluids}
\label{sec:evolution}

\begin{figure}
\includegraphics[width=0.6\textwidth]{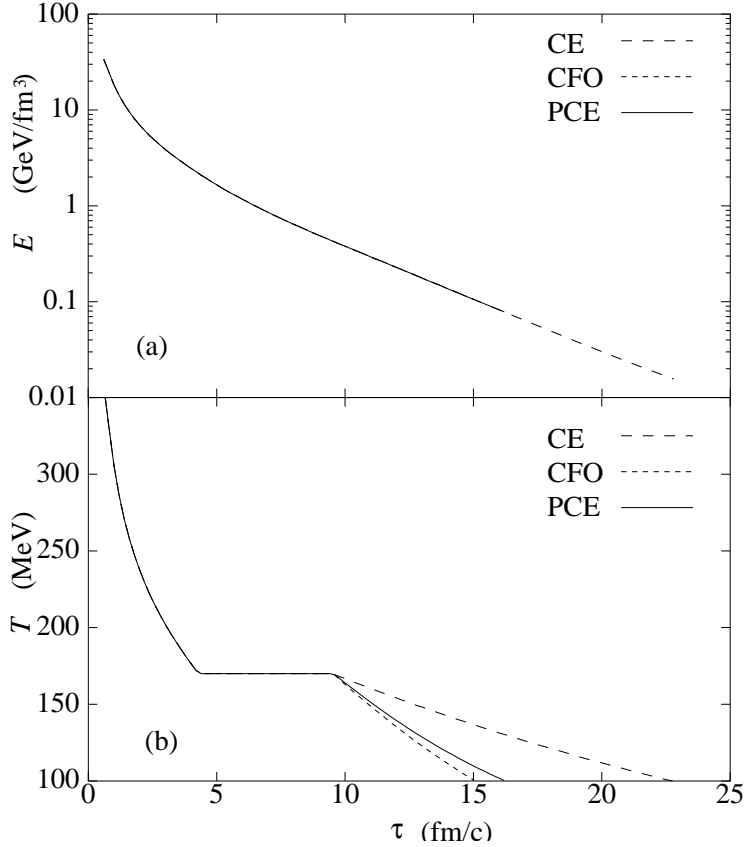}
\caption{\label{Fig3}(a) Time evolution of energy density at the center of the fluid. (b) Time evolution of temperature at the center of the fluid. The dashed, dotted, and solid lines correspond to the models CE, CFO, and PCE.}
\end{figure}

We simulate the space-time evolution of the fluid in the \textit{full} three dimensional space \cite{Hirano:2001eu} with the initial conditions and the EOS's discussed in the previous sections.
First, we pick up a fluid element at the central point ($x=y=\eta_{\mathrm{s}}=0$) and pursue its time evolution till its temperature reaches $T=100$ MeV.
Figure \ref{Fig3} shows the time evolution of (a) energy density and (b) temperature at the center of the fluid for three models EOS.
As far as the time evolution of energy density, we cannot distinguish each other.
This is easily understood by Fig.~\ref{Fig2} (a): Energy density evolution is completely governed by the EOS, i.e., $P(E)$ and the three models are very similar to each other.
We transform these results from energy density to temperature by using Fig.~\ref{Fig2} (b).
For the time evolution of temperature, chemical freeze-out makes a substantial difference between the model CE and the model CFO or PCE.
If we suppose thermal freeze-out occurs at the constant temperature, the system in which the property of the chemical freeze-out is considered has a fate to be thermally frozen earlier than the conventional model CE.
The early chemical freeze-out makes the hadron phase cool down more rapidly \cite{Arbex:2001vx}.

\begin{figure}
\includegraphics[width=0.6\textwidth]{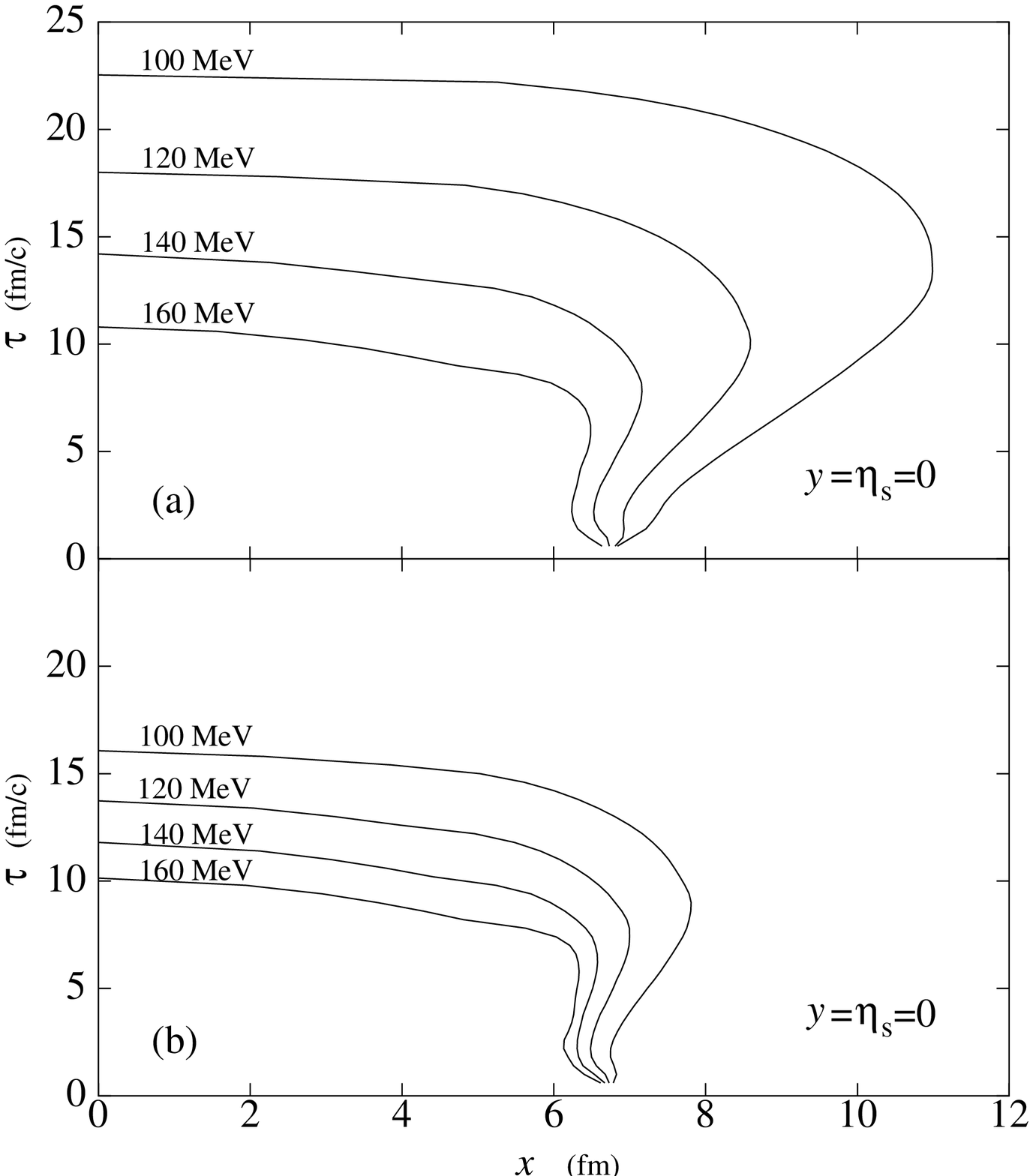}
\caption{\label{Fig4}Space-time evolution of thermal freeze-out hypersurface for (a) the model CE and (b) the model PCE.}
\end{figure}

Next, we show how the early chemical freeze-out affects the spatial size of the fluid.
Figure \ref{Fig4} represents the time evolution of freeze-out hypersurfaces at $y=\eta_{\mathrm{s}}=0$ for (a) the model CE and (b) the model PCE.
Here the hypersurfaces in Fig.~4 correspond to various $T^{\mathrm{th}} = 100, 120, 140$, and $160$ MeV which are within a plausible range for thermal freeze-out temperature.
We find that the early chemical freeze-out reduces not only the lifetime of the fluid but also its spatial size and that the fluid does not expand so explosively for the model PCE.
Since the two-particle correlation function is sensitive to the spatial size and the lifetime of the fluid, it is interesting to see the effect of early chemical freeze-out on the HBT radii. Detailed analyses of pion interferometry will be discussed in Sec.~\ref{sec:HBT}.

\begin{figure}
\includegraphics[width=0.6\textwidth]{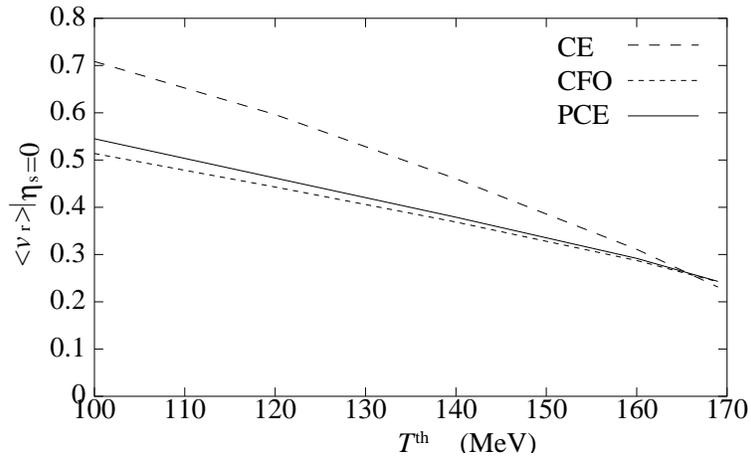}
\caption{\label{Fig5}Thermal freeze-out temperature dependence of average radial flow. The dashed, dotted, and solid lines correspond to the models CE, CFO, and PCE.}
\end{figure}

Figure \ref{Fig5} shows the thermal freeze-out temperature dependences of average radial flow at midrapidity $<v_r>\mid_{\eta_{\mathrm{s}} = 0}$, where $v_r = \sqrt{v_x^2 + v_y^2}$.
Radial flow is generated by the pressure gradient, so it contains informations about the EOS.
From this figure, the radial flow is suppressed when we take account of the early chemical freeze-out.
At $T^{\mathrm{th}}$ = 140 (120) MeV, the average radial flow for the model PCE is reduced by 17.7 (22.5) \% from the one for the conventional model CE.

It should be noted that the difference between the thermal freeze-out temperature $T^{\mathrm{th}}$ and the thermal freeze-out energy density $E^{\mathrm{th}}$ plays an important role in analyses of particle spectra. 
It may be claimed that there are no significant differences among three models when one shows the results in Figs.~\ref{Fig4} and \ref{Fig5} as functions of the thermal freeze-out \textit{energy density} $E^{\mathrm{th}}$, not temperature $T^{\mathrm{th}}$.
It is indeed true when one discusses only the hydrodynamic behavior.
The shape of particle distribution, especially $p_t$ spectrum, are determined by the thermal freeze-out \textit{temperature} $T^{\mathrm{th}}$ and the flow in the hydrodynamic model. So this difference is clearly meaningful when we compare the numerical results of particle spectra with experimental data as shown in the next section.

\section{Single Particle Spectra and Azimuthal asymmetry}
\label{sec:spectra}

Hydrodynamics has a limited prediction power, so the calculation is really meaningful only after tuning the initial parameters and reproducing the single particle spectra.
The momentum distribution for particles directly emitted from a (thermal) freeze-out hypersurface can be calculated through the Cooper-Frye formula \cite{Cooper:1974mv}
\begin{eqnarray}
\label{eq:CF}
E\frac{dN}{d^3 \bm{p}} = \frac{d}{(2 \pi)^3}\int_{\Sigma} \frac{p \cdot d\sigma}{\exp[(p \cdot u-\mu_i)/T^{\mathrm{th}}]\mp 1}.
\end{eqnarray}
Here, $\Sigma$ and $d\sigma^\mu$ are the thermal freeze-out hypersurface and its element. $u^\mu$ is the four fluid velocity. $-$ ($+$) sign is for boson (fermion) and $d$ is the degeneracy of particles under consideration. 
This formula merely counts the net particles passing through the hypersurface $\Sigma$ rather than decoupling from the system.
Although it has a problem on the negative number in the treatment of time-like freeze-out hypersurface \cite{freezeout}, this is widely used in almost all hydrodynamic models.
The observed spectra always contain the contribution from resonance decays.
We assume all of the resonance particles in the EOS are also emitted from a freeze-out hypersurface and that they decay into stable particles.
Taking account of the decay kinematics, we easily obtain the single particle spectra from resonance particles \cite{Sollfrank:1991xm,Hirano:2000eu}.\footnote{When we calculate the two-pion correlation function in the next section, we neglect this contribution for simplicity.}
For further details to calculate the spectra from resonance decays, see Appendix.
Since the results from the model CFO is similar to the ones from the model PCE, we hereafter concentrate our discussions on the models CE and PCE.

\subsection{Spectra and Flow for Charged Hadrons}

\begin{figure}
\includegraphics[width=0.6\textwidth]{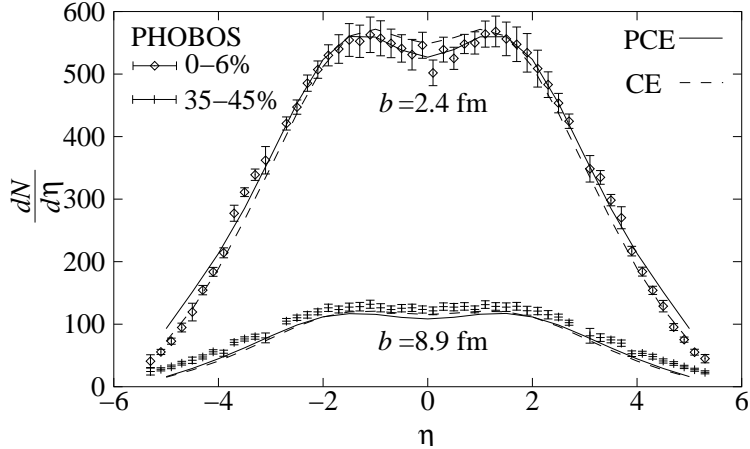}
\caption{\label{Fig6}Pseudorapidity distribution of charged particles in Au+Au 130$A$ GeV central and semi-central collisions. Data from Ref.~\cite{Back:2001bq}.}
\end{figure}

Figure \ref{Fig6} shows the pseudorapidity distribution of charged particles in Au + Au 130$A$ GeV collisions.
We choose $T^{\mathrm{th}}$ = 140 MeV for both models.
From the analyses based on the wounded nucleon model, we choose the impact parameter $b=2.4$ fm for 0-6 \% central collisions and $b=8.9$ fm for 35-45 \% non-central collisions.
The resultant number of participants is 342 (94) for $b=2.4$ (8.9) fm, which is consistent with estimation by the PHOBOS Collaboration \cite{Back:2001bq}.
We reasonably reproduce the data in not only central but also non-central collisions by using initial parameters in the previous section.
After tuning initial parameters for central events, we have no adjustable parameters for non-central events due to assuming the binary collision scaling. For non-central collisions, we change the impact parameter in Eq.~(\ref{eq:Tprofile}) as to the number of participants with keeping other parameters in the initial condition.
Although the binary collisions contribute to hard components, the binary collision scaling seems to be reasonable to parameterize the hydrodynamic initial condition \cite{Kolb:2001qz} from Fig.~\ref{Fig6}.

\begin{figure}
\includegraphics[width=0.6\textwidth]{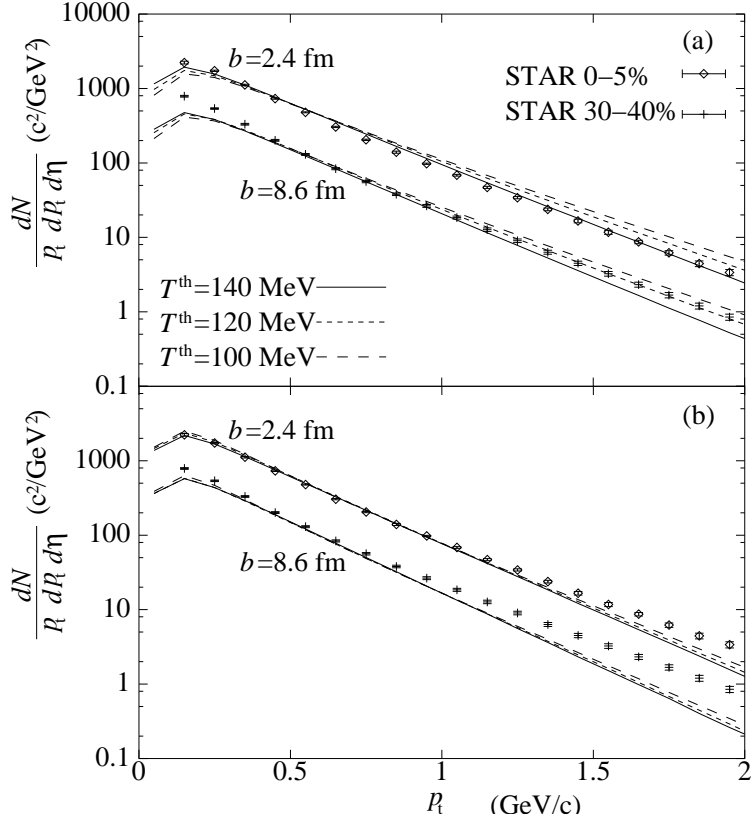}
\caption{\label{Fig7}Transverse momentum spectra of negative charged particles in Au+Au 130$A$ GeV central and semi-central collisions for (a) the model CE and (b) the model PCE. The dashed, dotted, and solid lines correspond to $T^{\mathrm{th}}$ = 100, 120, and 140 MeV. Data from Refs.~\cite{Adler:2001yq} and \cite{CalderondelaBarcaSanchez:2001np}.}
\end{figure}

We next show in Fig.~\ref{Fig7} the $p_t$ spectrum for negative charged hadrons.
The experimental data (0-5\% central \cite{Adler:2001yq} and 30-40 \% semi central events \cite{CalderondelaBarcaSanchez:2001np}) are observed by the STAR Collaboration.
The impact parameters used in this calculation are $b=2.4$ fm for central and $b=8.6$ fm for semi central events.
We represent the thermal freeze-out temperature dependence of $p_t$ spectrum for the models CE and PCE.
The slope of $p_t$ spectrum is almost independent of $T^{\mathrm{th}}$ near $p_t \sim 1$ GeV/$c$, which is a little peculiar behavior in the usual sense.
However, it is interpreted by the result in Fig.~\ref{Fig5}.
When one reduces the thermal freeze-out temperature by hand, the average radial flow enhances as its response.
The magnitude of the response is governed by the EOS.
The resultant $p_t$ slope is a competition between these two effects:
The reduction of temperature makes the slope steeper in the case of vanishing flow, while the thermal distribution is Lorentz-boosted by radial flow and the $p_t$ slope becomes flatter.
The effect of generated radial flow on the $p_t$ slope usually overcomes that of the reduction of $T^{\mathrm{th}}$ in the model CE, so the $p_t$ slope becomes flatter as decreasing $T^{\mathrm{th}}$.
On the other hand, the radial flow is slightly suppressed in the model PCE as shown in Fig.~\ref{Fig5}.
Hence the reduction of $T^{\mathrm{th}}$ is just compensated by its response to radial flow for the model PCE.
This is the reason why the $p_t$ slope is almost independent of $T^{\mathrm{th}}$ in Fig.~\ref{Fig7}.
For the model CE, we reproduce the slope by choosing $T^{\mathrm{th}}$ = 140 MeV.
For the model PCE, in any $T^{\mathrm{th}}$ within a plausible range, we reproduce the experimental data below 1 GeV/$c$ for central collisions.
This indicates that there exists the onset of hard processes around $p_t \sim 1$ GeV/$c$.
It should be noted that a bend of the spectrum in low $p_t$ region is simply due to the Jacobian between the pseudorapidity $\eta$ and the rapidity $Y$.

\begin{figure}
\includegraphics[width=0.6\textwidth]{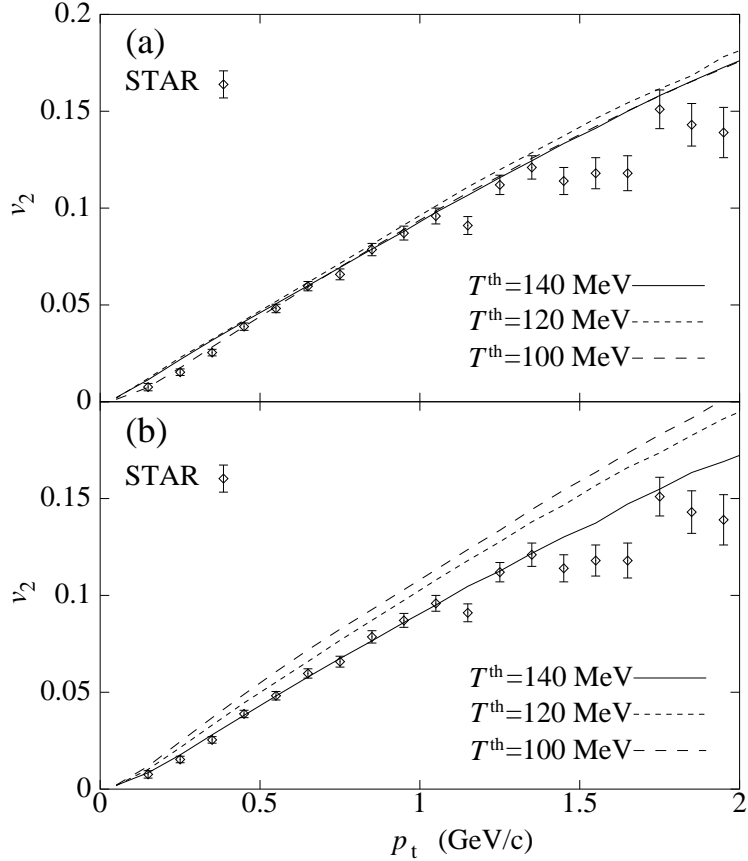}
\caption{\label{Fig8}$v_2(p_t)$ for charged hadrons in Au+Au 130$A$ GeV collisions. The dashed, dotted, and solid lines correspond to $T^{\mathrm{th}}$ = 100, 120, and 140 MeV. Data from Ref.~\cite{Ackermann:2000tr}.}
\end{figure}

We next show the transverse momentum dependence of the second Fourier coefficient for azimuthal distribution for the models CE and PCE.
Figure \ref{Fig8} represents $v_2(p_t)$ for charged particles in minimum bias collisions.
Experimental data is also observed by STAR \cite{Ackermann:2000tr}.
Hydrodynamic analysis of the data has already been done by Kolb \text{et al.} \cite{Kolb:2000fh}. They found the hydrodynamic result excellently coincides with the data below $p_t \sim 1.5$ GeV/$c$.
Our results for the model CE are consistent with their results.
Now we find the space-time evolution in our model is different from their result where chemical equilibrium is always assumed, we must check whether the hydrodynamic description is really good at RHIC even when we include the effects of early chemical freeze-out.
The numerical results are calculated from the following equation:
\begin{eqnarray}
v_2(p_t) & = & \frac{\sum_b \int d\eta d\phi \cos(2\phi) b \frac{dN}{p_t dp_t d\eta d\phi}(p_t, \eta, \phi; b)}{\sum_b \int d\eta \enskip b \frac{dN}{p_t dp_t d\eta}(p_t, \eta; b)}.
\end{eqnarray}
Here the summation with respect to the impact parameter $b$ is taken over every 2 fm up to 14 fm in this analysis. The integral region of $\eta$ is from $-1.3$ to $1.3$, which corresponds to the analysis by STAR \cite{Ackermann:2000tr}.
The value of $v_2(p_t)$ depends on the thermal freeze-out temperature $T^{\mathrm{th}}$ in contrast with the $p_t$ spectrum.
We also reproduce the experimental data below 1 GeV/$c$ by choosing $T^{\mathrm{th}} = 140$ MeV and slightly overestimate the data above 1 GeV/$c$.
Similar to the $p_t$ spectrum, this result also indicates that the hard contribution, which reduces $v_2$ calculated from hydrodynamic source \cite{Gyulassy:2001kr,Wang}, becomes important above 1 GeV/$c$.

\begin{figure}
\includegraphics[width=0.6\textwidth]{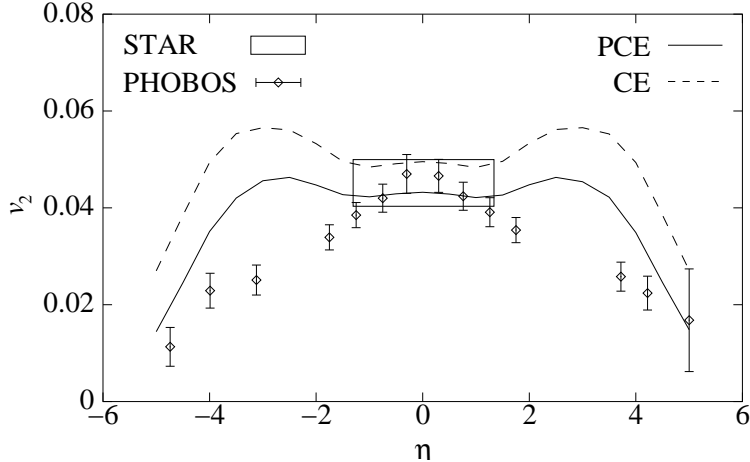}
\caption{\label{Fig9}$v_2(\eta)$ for charged hadrons in Au+Au 130$A$ GeV collisions. The solid and dashed lines correspond to the models PCE and CE. Data from Refs.~\cite{Ackermann:2000tr} and \cite{Park:2001gm}.}
\end{figure}

In Fig.~\ref{Fig9}, we compare the pseudorapidity dependence of elliptic flow between the models CE and PCE.
Similar to the $v_2(p_t)$, $v_2(\eta)$ which is to be compared with minimum bias data is
\begin{eqnarray}
v_2(\eta) & = & \frac{\sum_b \int p_t dp_t d\phi \cos(2\phi) b \frac{dN}{p_t dp_t d\eta d\phi}(p_t, \eta, \phi; b)}{\sum_b \int p_t dp_t \enskip b \frac{dN}{p_t dp_t d\eta}(p_t, \eta; b)}.
\end{eqnarray}
We choose $T^{\mathrm{th}} = 140$ MeV and integrate with respect to $p_t$ from 0 to 2.0 GeV/$c$. 
Data plots are observed by PHOBOS \cite{Park:2001gm}.
The rectangular area corresponds to the statement by STAR \cite{Ackermann:2000tr}, $v_2(\eta) = 4.5 \pm 0.5 \%$ for $0.1 < p_t < 2.0$ GeV/$c$ and $\mid \eta \mid < 1.3$.
As well as the case of radial flow, the elliptic flow is also reduced by taking account of the chemical freeze-out.
We reproduce the PHOBOS data only near mid(pseudo)rapidity and overestimate in forward and backward rapidity.
On the other hand, a microscopic transport model (JAM) reproduces the data only in forward and backward rapidity regions \cite{JAM}.
This indicates that the full thermalization is achieved only near midrapidity, although there are some open problems in hydrodynamics such as the treatment of freeze-out through the Cooper-Frye formula, more sophisticated initialization, and the absorption by spectators \cite{Hirano:2001eu}.

From Figs.~\ref{Fig8} and \ref{Fig9}, the hydrodynamic description with early chemical freeze-out seems to be valid for, at least, $0< p_t < 1$ GeV/$c$ and $-1 < \eta < 1$ in Au+Au collisions at $130A$ GeV.

\subsection{Spectra and Flow for Identified Hadrons}

In this subsection, let us see the difference between the models CE and PCE by comparing $p_t$ spectra and $v_2(p_t)$ for $identified$ hadrons which are supposed to be sensitive to the early chemical freeze out.

\begin{figure}
\includegraphics[width=0.6\textwidth]{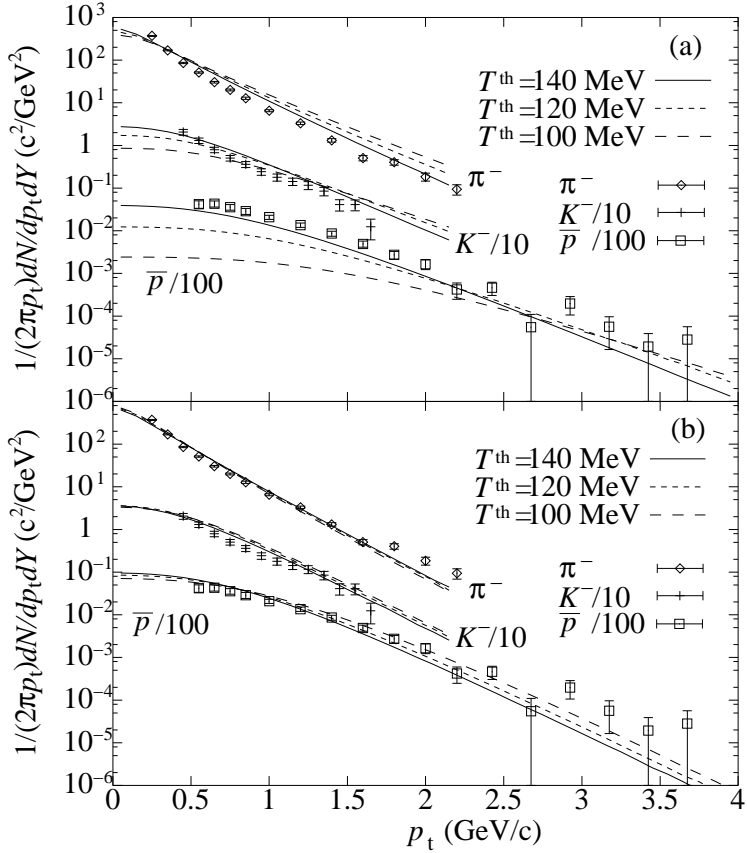}
\caption{\label{Fig10} The transverse momentum spectra for negative pions, negative kaons, and antiprotons for (a) the model CE and (b) the model PCE.
To see these results clearly, the yield of kaons (antiprotons) is scaled by 10$^{-1}$ (10$^{-2}$).
The dashed, dotted, solid lines represent results for $T^{\mathrm{th}}=$ 100, 120, and 140 MeV, respectively.
Data from Ref.~\cite{Adcox:2001mf}.}
\end{figure}

The $p_t$ spectra for identified hadrons in (0-5\%) central collisions observed by the PHENIX Collaboration \cite{Adcox:2001mf} are compared with our results in Fig.\ref{Fig10}.
The impact parameter which we choose for central collisions is also $b=2.4$ fm.
For the model CE, the slopes of pions, kaons, and antiprotons become steeper as increasing $T^{\mathrm{th}}$, which is similar to the case of charged particles.
The number of antiproton becomes very small at $T^{\mathrm{th}} = 100$ MeV for the model CE due to chemical equilibrium.
On the other hand, the numbers of each hadron in the model PCE are independent of $T^{\mathrm{th}}$ and reasonably agree with experimental data.
The number of anti-proton in the model PCE might be slightly improved by taking into account the baryonic chemical potential which is neglected in the present analysis.
It should be noted that pion spectra in large $p_t$ ($>$1.5 GeV/$c$) region can be reproduced by including contribution from non-thermalized hard partons (jets) with energy loss \cite{hydrojet}.

\begin{figure}
\includegraphics[width=0.8\textwidth]{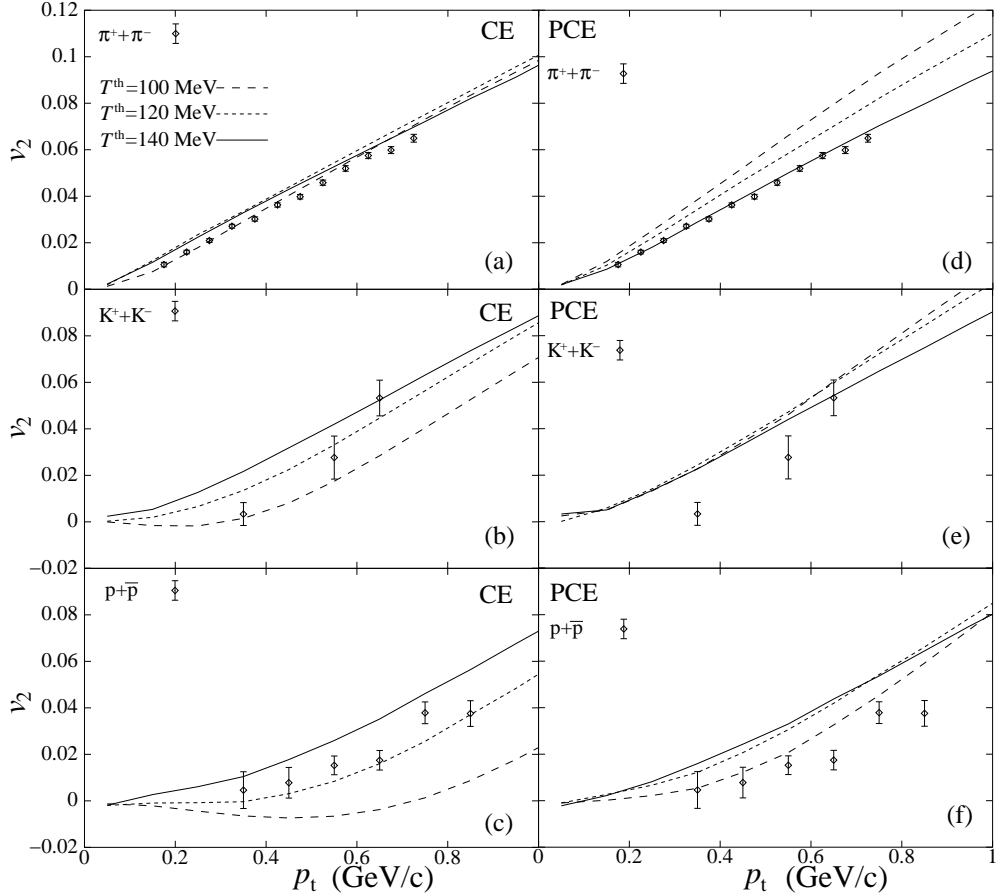}
\caption{\label{Fig11} $v_2(p_t)$ for pions, kaons, and protons in Au + Au 130$A$ GeV collisions. Left (right) column represents the results for the model CE (PCE). Data from Ref.~\cite{Adler:2001nb}.}
\end{figure}

We next show in Fig.~\ref{Fig11} the elliptic flow for identified hadrons and its $T^{\mathrm{th}}$ dependence.
The STAR data \cite{Adler:2001nb} are compared with our results with or without chemical equilibrium.
For the model CE, $v_2(p_t)$ of pions is almost independent of $T^{\mathrm{th}}$, while $v_2(p_t)$ of kaons and protons are increasing with $T^{\mathrm{th}}$.
On the other hand, $v_2$ of pions grows with decreasing $T^{\mathrm{th}}$ for the model PCE.
Whether $v_2(p_t)$ increases with $T^{\mathrm{th}}$ depends not only on the particle mass but also on the flow velocities and its anisotropy in the transverse plane \cite{Huovinen}.
The elliptic flow seems to be more sensitive to $T^{\mathrm{th}}$ than $p_t$ spectra when we consider the early chemical freeze-out.
For the transverse momentum spectra of identified hadron shown in Fig.~\ref{Fig7}, we roughly reproduce the slope in low $p_t$ region with $T^{\mathrm{th}}$ = 140 MeV. On the other hand, we cannot reproduce $v_2(p_t)$ of identified hadrons by a common thermal freeze-out temperature: each hadron seems to favor different $T^{\mathrm{th}}$. This indicates the hadronic afterburner in the late stage of the expansion may be important \cite{RQMD}. From hydrodynamic point of view, detailed analyses with various EOS's and initial conditions are needed in understanding thermal freeze-out properties.

\begin{figure}
\includegraphics[width=0.6\textwidth]{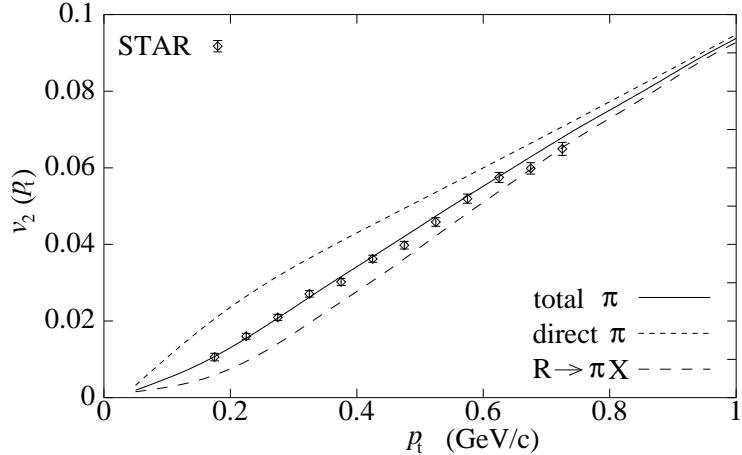}
\caption{\label{Fig12} $v_2(p_t)$ for charged pions. The solid, dotted, and dashed lines correspond to total pions, pions directly emitted from freeze-out hypersurface, and pions from resonance decays. Data from Ref.~\cite{Adler:2001nb}.}
\end{figure}

Figure \ref{Fig12} separately represents the contribution of pions directly emitted from freeze-out hypersurface and the contribution of pions from resonance decays.
The STAR data shows the contribution from identified pions \cite{Adler:2001nb}.
From this figure, the concavity of $v_2(p_t)$ in low $p_t$ region ($p_t<0.3$ GeV/$c$) results from the resonance decay after thermal freeze-out.
The exact treatment of the decay kinematics of resonances dilutes $v_2$ for direct pions especially in low $p_t$ region \cite{Hirano:2000eu}.
The fraction of the contribution from resonance decays plays a very important role in understanding $v_2(p_t)$ for pions in low $p_t$ region, so the early chemical freeze-out must be included when we discuss proper phenomena to the low transverse momentum.

\section{PION INTERFEROMETRY}
\label{sec:HBT}

From the single particle spectra and the azimuthal asymmetry, we obtain the information about the distribution in the momentum space at freeze-out. On the other hand, we can obtain the information about the particle distribution in the coordinate space through two-particle interferometry \cite{HBT}. 
We see in Sec.~\ref{sec:evolution} the space-time evolutions are considerably different between the models with and without chemical equilibrium. To see this more clearly, we discuss the two-pion correlation function in this section.

In hydrodynamics, the two-particle correlation function for directly emitted bosons from freeze-out hypersurface $\Sigma$ can be calculated from \cite{Schlei:1992jj,Rischke:1996em}
\begin{eqnarray}
C_2(\bm{p}_1,\bm{p}_2) & = & \frac{P(\bm{p}_1, \bm{p}_2)}{P(\bm{p}_1)P(\bm{p}_2)} \nonumber \\
& = & 1+\frac{\left| \frac{d}{(2\pi)^3} \int_{\Sigma} K \cdot d\sigma  \exp(i x \cdot q) f_{\mathrm{BE}}(\frac{K \cdot u -\mu_i}{T^{\mathrm{th}}}) \right|^2}{E_1 \frac{dN}{d^3\bm{p}_1} E_2\frac{dN}{d^3\bm{p}_2}}.
\end{eqnarray}
Here $P(\bm{p}_1, \bm{p}_2)$ is the two particle coincidence cross section and $P(\bm{p}_1)$ is the same Cooper-Frye formula represented in Eq.~(\ref{eq:CF}).
We consider only directly emitted pions for simplicity.
$K^\mu = (p_{1}^{\mu}+p_{2}^{\mu})/2$ and $q^\mu = p_{1}^{\mu}-p_{2}^{\mu}$ are, respectively, the average and relative four momentum of pair. $f_{\mathrm{BE}}=(e^x-1)^{-1}$ is the Bose-Einstein distribution function.
The informations about hydrodynamic simulations enter through the freeze-out hypersurface $\Sigma$ and the four velocity $u^\mu$ in this equation.
The average pair momentum $K$ is decomposed into the transverse momentum $K_T$, the longitudinal momentum $K_z$, and the azimuthal angle $\Phi_K$.\footnote{Even for central events, we have no longer cylindrical symmetry around the collision axis due to small (but finite) value of the impact parameter, so $C_2$ depends on the azimuthal angle $\Phi_K$ which is measured from the reaction plane.}
The relative pair momentum $q^\mu$ is also decomposed into the standard coordinate, $q_{\mathrm{out}}$ (parallel to $K_T$), $q_{\mathrm{long}}$ (along the beam direction), and $q_{\mathrm{side}}$ (perpendicular to the others).
Since the experimental acceptances are limited to midrapidity, $\mid Y \mid <0.5$ (STAR) \cite{Adler:2001zd} or $\mid \eta \mid < 0.35$ (PHENIX) \cite{Adcox:2002uc}, we can put $K_z = 0$.
Moreover, we average the two-particle function $C_2$ over the azimuthal angle $\Phi_K$.
Thus we obtain the following equation which can be compared with the experimental data:
\begin{eqnarray}
C_2(K_T, q_{\mathrm{side}}, q_{\mathrm{out}}, q_{\mathrm{long}})  =  \left.\frac{\int \Phi_K P(\bm{p}_1, \bm{p}_2)}{\int \Phi_K P(\bm{p}_1)P(\bm{p}_2)}\right|_{K_z = 0}
\end{eqnarray}
Our definition of the HBT radii is similar to the one in Refs.~\cite{Rischke:1996em,Soff:2000eh,Zschiesche:2001dx}. Assuming that $C_2(K_T, q_{\mathrm{side}}, q_{\mathrm{out}}, q_{\mathrm{long}})$ for each $K_T$ is fitted by the Gaussian form 
\begin{eqnarray}
C_2 = 1 + \lambda \exp(-R_{\mathrm{side}}^2 q_{\mathrm{side}}^2-R_{\mathrm{out}}^2 q_{\mathrm{out}}^2-R_{\mathrm{long}}^2 q_{\mathrm{long}}^2)
\end{eqnarray}
with a chaoticity $\lambda = 1$, the $K_T$ dependence of HBT radius for the side direction is $R_{\mathrm{side}}(K_T) = 1/q_{\mathrm{side}}^* (K_T)$, where $C_2(K_T, q_{\mathrm{side}}^*,0,0) = 1+e^{-1}$, and analogous definitions for the $K_T$ dependence of $R_{\mathrm{out}}$ and $R_{\mathrm{long}}$.

\begin{figure}
\includegraphics[width=0.6\textwidth]{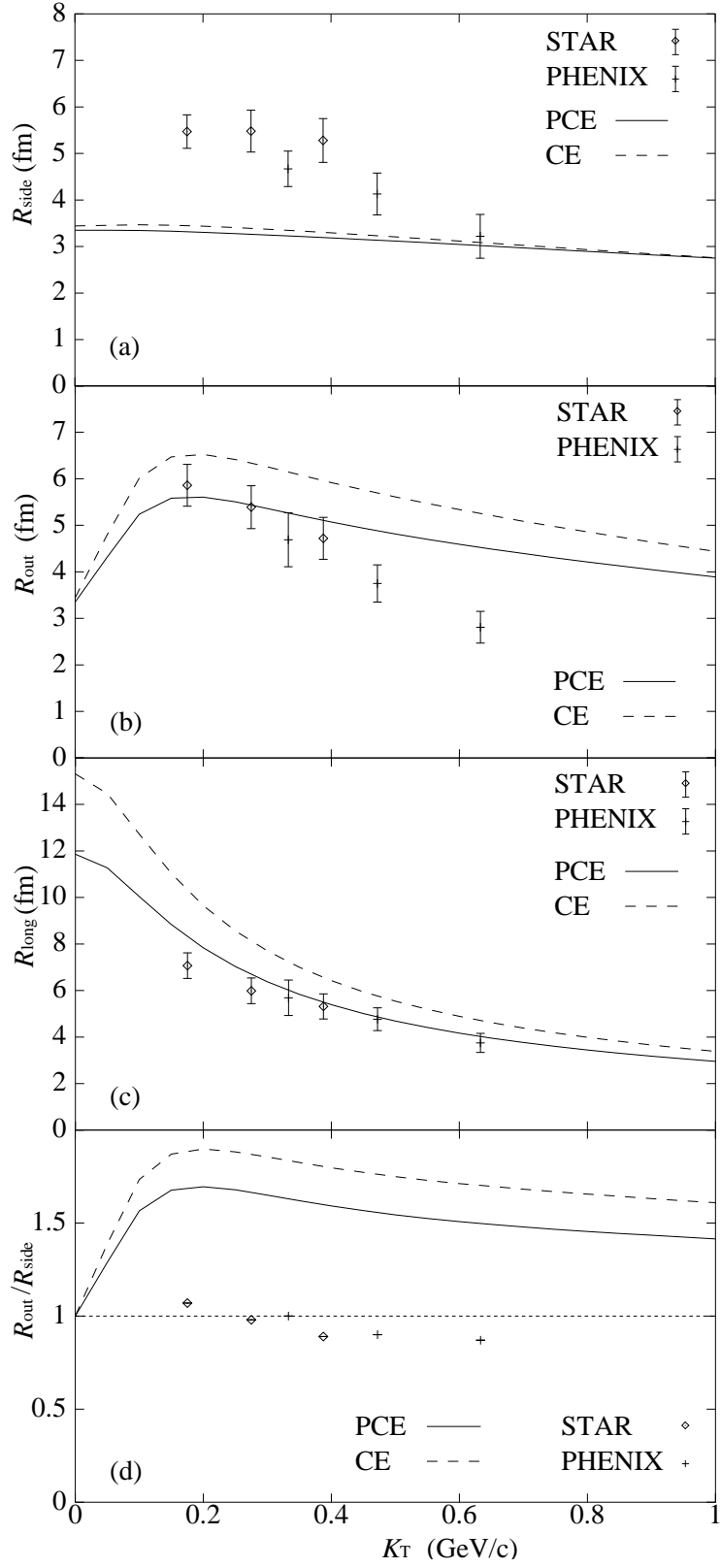}
\caption{\label{Fig13} HBT radii for negative pions in (a) side, (b) out, and (c) long direction and (d) its ratio  $R_{\mathrm{out}}/R_{\mathrm{side}}$ as a function of $K_T$. The solid and dashed lines correspond to the models PCE and CE. Data from Refs.~\cite{Adler:2001zd} and \cite{Adcox:2002uc}.}
\end{figure}

We evaluate two-pion correlation functions for negative pions directly emitted from freeze-out hypersurface and obtain the $K_T$ dependence of $R_{\mathrm{side}}$, $R_{\mathrm{out}}$, and $R_{\mathrm{long}}$.
Figure \ref{Fig13} shows the HBT radii and the ratio $R_{\mathrm{out}}/R_{\mathrm{side}}$ for the models CE and PCE with $T^{\mathrm{th}} = 140$ MeV.
Here the impact parameter which we choose in this analysis is $b=2.4$ fm.
The difference between the two models is very small for $R_{\mathrm{side}}$.
On the other hand, $R_{\mathrm{out}}$ and $R_{\mathrm{long}}$ for the model PCE are significantly smaller than the ones for the model CE, which reflects the space-time evolution of freeze-out hypersurface depicted in Fig.~\ref{Fig4}.
We compare our results with experimental data observed by STAR (12\% most central events) \cite{Adler:2001zd} and PHENIX (30\% most central events) \cite{Adcox:2002uc}.
We reproduce the $K_T$ dependence of $R_{\mathrm{long}}$ by employing the model PCE, while $R_{\mathrm{side}}$ and $R_{\mathrm{out}}$ do not show good coincidences with the experimental data.

It has been suggested that the ratio $R_{\mathrm{out}}/R_{\mathrm{side}}$ reflects the prolongation of the lifetime due to the phase transition between the QGP phase and the hadron phase \cite{Rischke:1996em}.
Various models predict this ratio has a value significantly larger than unity in some $K_T$ region \cite{Hirano:2001yi,Rischke:1996em,Soff:2000eh,Zschiesche:2001dx,Heinz}, although it is around unity in $0.2< K_T < 0.6$ GeV/$c$ in Au + Au $\sqrt{s_{NN}} = 130A$ GeV collisions according to the recent measurement at RHIC \cite{Adler:2001zd,Adcox:2002uc}.
This discrepancy is often called ``the HBT puzzle" \cite{Heinz,Soff:2002qw}.
The ratio for the model PCE reduces by about 12 \% above 0.2 GeV/$c$ due to the property of the early chemical freeze-out, but it clearly turns out to be larger than the experimental data.
Our aim in this paper is to see how the early chemical freeze-out affects the HBT radii, so we leave detailed discussions on the HBT puzzle in the future works.\footnote{
In Ref.~\cite{Hirano:2001yi}, slightly larger $R_{\mathrm{side}}$ is obtained by assuming simple flat transverse profile and vanishing impact parameter for initial energy density. The flat profile leads a larger value of variance of a fluid in the transverse plane than the present initialization based on the binary collision scaling.
}

\section{Summary and Discussion}
\label{sec:summary}

We have investigated the effect of early chemical freeze-out on the hydrodynamic evolution and the particle spectra by using a genuine three dimensional hydrodynamic model.
We constructed the equation of state for hadronic matter which is in partial chemical equilibrium.
Pressure as a function of energy density is not affected by the chemical non-equilibrium property, while temperature as a function of energy density is largely reduced due to the large population of resonance particles.
By using the EOS with a first order phase transition, we simulated hydrodynamic evolution at the RHIC energy.
We found the system cools down more rapidly than the conventional model
and that the lifetime and the spatial size of the fluid and the radial and elliptic flows are reduced when we compared at isothermal hypersurface.
We also analyzed particle spectra and two-pion correlation functions in Au+Au $130A$ GeV collisions. We chose initial parameters in the hydrodynamic simulations so as to reproduce the pseudorapidity distribution in central collisions observed by PHOBOS. 
The slope of $p_t$ spectrum for negative hadrons is less sensitive to the thermal freeze-out temperature, which results from the reduction of radial flow.
On the other hand, the transverse momentum dependence of elliptic flow $v_2(p_t)$ for charged hadrons depends on the thermal freeze-out temperature.
The situation is completely opposite to the ordinary hydrodynamic results. In the conventional models, the $p_t$ slope is steeper as increasing $T^{\mathrm{th}}$ and $v_2(p_t)$ is not so sensitive to $T^{\mathrm{th}}$.
We found $T^{\mathrm{th}}=140$ MeV and the resultant average radial flow $<v_r> = 0.38c$ are the values to simultaneously reproduce the $p_t$ slope and $v_2(p_t)$ for charged hadrons (mainly pions) below 1 GeV.
$v_2(\eta)$ is also reduced by the early chemical freeze-out, but we failed to reproduce the data in forward and backward rapidity regions. 
We see the thermal freeze-out temperature independence of $p_t$ slope more clearly in the $p_t$ spectra for identified hadrons. We reasonably reproduce the $p_t$ slope with $T^{\mathrm{th}}=100$-$140$ MeV for the model PCE, while $v_2(p_t)$ for identified hadrons seems to favor different thermal freeze-out temperatures which are dependent on hadronic species.
In order to see more quantitatively the effect of early chemical freeze-out on the temporal or spatial size of fluids, we calculated the two-pion correlation functions and obtained the HBT radii.
By taking account of the early chemical freeze-out in the EOS, $R_{\mathrm{out}}$, $R_{\mathrm{long}}$, and $R_{\mathrm{out}}/R_{\mathrm{side}}$ are significantly suppressed, while $R_{\mathrm{side}}$ is not changed.
Nevertheless, the properties of the early chemical freeze-out are not enough to interpret the HBT puzzle.

It would be very interesting to see other observables in the case of partial chemical equilibrium.
Although the best place to see the difference between the models CE and PCE must be the particle ratios, we cannot discuss this observable by using the present model due to an approximation of vanishing baryonic chemical potential.
This will be discussed elsewhere.
Penetrating probes such as thermal photons and dileptons may also be affected by the early chemical freeze-out.
The emission rates of photons or dileptons increase due to the chemical potentials for hadrons, while the space-time volume of the hadron phase decreases.
Therefore we should check whether the total multiplicity and spectra of photons and dileptons are changed in terms of hydrodynamics.

There are similar dynamical approaches to describe the early chemical freeze-out. 
The model evolves the QGP and mixed phases as a relativistic fluid, while it switches to a hadronic cascade model, e.g., UrQMD \cite{URQMD} or RQMD \cite{RQMD}.
The advantage of our hydrodynamic model over these hybrid (hydrodynamics + cascade) models is to be able to obtain the average hydrodynamic behavior and the effect of temperature naturally.
When one discusses the spectral changes of hadrons due to the medium effects such as temperature and/or baryonic chemical potential \cite{HATSUDA}, one can easily estimate its effect by using hydrodynamic simulations.
We regard the model PCE as a complementary tool to those hybrid models.

After publishing our preliminary results \cite{Hirano:2002qj} and almost finishing this work, the authors are aware of a paper concerning the same subject \cite{Teaney:2002aj}, in which the conclusion on the hydrodynamic behavior is almost the same as ours. In Ref.~\cite{Teaney:2002aj}, the effect of finite baryonic chemical potential is included but the hydrodynamic simulation is performed only in the transverse plane by assuming the Bjorken's scaling solution.

\begin{acknowledgments}
The authors thank members in high-energy physics group at Waseda University for useful comments and K.~Kajimoto for his collaboration in the early stage of this work.
One of the author (T.H.) acknowledges valuable discussion with T.~Hatsuda, P.~Huovinen, K.~Morita, S.~Muroya, H.~Nakamura, C.~Nonaka, and D.~Rischke.
He also thanks Inoue Foundation for Science for financial support.
\end{acknowledgments}

\appendix*
\section{MONTE CARLO CALCULATION OF THE CONTRIBUTION FROM RESONANCE DECAYS}
\label{MC}

In this appendix, we show how to calculate the particle distribution from resonance decays within the Cooper-Frye prescription.
This method can be used in hydrodynamics-motivated models as well as in hydrodynamics.

Lorentz transformation for the momentum of a decay particle between the local rest system (starred) and the finite momentum system (non-starred) of a resonance particle $R$ is 
\begin{eqnarray}
\label{LOTR}
\bm{p}^* = \bm{p}-\bm{p}_{R}\left[\frac{E}{m_{R}}-\frac{\bm{p}\cdot \bm{p}_{R}}{m_{R}(m_{R}+E_{R})} \right].
\end{eqnarray}
We rewrite Eq.~(\ref{LOTR}) explicitly 
\begin{eqnarray}
p^*_l &=& p_l-p_{Rl}F(p_l,\phi), \\
\cos \phi^* &=& \frac{p_x^*}{p_t^*} \nonumber\\
            &=& \frac{p_t(p_l, \phi) \cos \phi-p_{Rt}\cos \phi_{R}F(p_l,\phi)}{\sqrt{p_t^2(p_l, \phi)+p_{Rt}^2 F^2(p_l,\phi)-2p_t(p_l, \phi)p_{Rt}\cos(\phi-\phi_{R})F(p_l,\phi)}},
\end{eqnarray}
where
\begin{eqnarray}
F(p_l,\phi) = \frac{E(p_l, \phi)}{m_{R}}-\frac{p_t(p_l,\phi) p_{Rt}\cos(\phi-\phi_{R})+p_l p_{Rl}}{m_{R}(m_{R}+E_{R})}.
\end{eqnarray}
Here the independent variables which we choose for decay particles are the longitudinal momentum $p_l$ and the azimuthal angle $\phi$.
Thus the transverse momentum of a decay particle $p_t$ is written in terms of $p_l$ and $\phi$
\begin{eqnarray}
p_t(p_l, \phi) & = & \frac{1}{\gamma_{R}\left[1-v_{Rt}^2\cos^2(\phi-\phi_{R})
\right]}
\bigg\{(E^* + p_l v_{Rl} \gamma_{R})v_{Rt}\cos(\phi-\phi_{R}) \nonumber \\
& \pm & \left.\sqrt{(E^* + p_l v_{Rl} \gamma_{R})^2-(p_l^2+m^2)\gamma_{R}^2\left[1-v_{Rt}^2\cos^2(\phi-\phi_{R})\right]} \right\}.
\end{eqnarray}
The Jacobian of the Lorentz transformation is defined by
\begin{eqnarray}
dp_l^*d\phi^* & = & J(p_l,\phi;\bm{V}_{R})dp_ld\phi,\\
\label{JACOBI}
J(p_l,\phi;\bm{V}_{R}) & = & \left|
 \begin{array}{cc}
  \frac{\partial p_l^*}{\partial p_l} & \frac{\partial p_l^*}{\partial \phi}\\
  \frac{\partial \phi^*}{\partial p_l} & \frac{\partial \phi^*}{\partial \phi}
 \end{array}
\right|.
\end{eqnarray}
The calculation of $J$ is straightforward, so that we do not represent it here.
The normalization of momentum space volume for a decay particle in the resonance rest frame is 
\begin{eqnarray}
\int_{-p^*}^{p^*}\frac{dp_l^*}{2p^*} \int_0^{2 \pi} \frac{d\phi^*}{2 \pi} = 1.
\end{eqnarray}
We always average the decay probability over the spin of resonances, so that the decay probability does not depend on $p_l^*$ and $\phi^*$.
Thus the normalization in the resonance reference frame is
\begin{eqnarray}
\int\frac{J(p_l,\phi;\bm{V}_{R})dp_ld\phi}{4\pi p^*} = 1. 
\end{eqnarray}

The Jacobian in Eq.~(\ref{JACOBI}) has very narrow peaks when the resonance particle moves at a large velocity in the laboratory system \cite{Hirano:2000eu}. This singularity makes it difficult to integrate the Jacobian numerically. So we introduce a very simple Monte Carlo calculation to evaluate the momentum distribution from resonance decays.
All input parameters in this calculation are the numerical results of hydrodynamic simulation, i.e., the temperature $T^{\mathrm{th}}$, the chemical potential for resonance particles $\mu_R$, the three-dimensional fluid velocity $\bm{v}$, and the element $d \sigma^\mu$ on the freeze-out hypersurface $\Sigma$.
In the following discussion, we show how to obtain the rapidity distribution of negative pions, for simplicity, only from $\rho$ mesons.
In this case, the branching ratio $B_{\rho^{0(-)} \rightarrow \pi^- \pi^{+(0)}} = 1$.
It is straightforward to extend this scheme to the cases for other resonances or the transverse mass (momentum) distribution.

\noindent
\textit{Step 1}: Evaluate the number of $\rho^0$ and $\rho^-$ which are emitted from \textit{or} absorbed by the $k$-th freeze-out hypersurface element $d \sigma^\mu_k$:
\begin{eqnarray}
N^{R}_k = \frac{g_{R}}{(2 \pi)^3}\int_{\Sigma} \frac{d^3p_{R}}{E_{R}} \frac{\mid p_{R} \cdot d\sigma_k \mid}{\exp\left[\left(p_{R} \cdot u_k -\mu_R \right)/T^{\mathrm{th}}\right] - 1}.
\end{eqnarray}
The integrand does not contain the Jacobian, so that it is simple to carry out the numerical integration by a standard technique.
It should be noted that $N^{R}_k$ is different from the \textit{net} number of emitted $\rho$ mesons from the $k$-th fluid element and that this is used merely for normalization.

\noindent
\textit{Step 2}: Generate $\tilde{N}$ random momenta $P^*_j$ ($1\le j \le \tilde{N}$) for $\rho$ mesons which obey the distribution
\begin{eqnarray}
\frac{{P^*}^2}{\exp\left[\left(\sqrt{{P^*}^2+m^2_{R}} - \mu_R\right)/T^{\mathrm{th}}\right] - 1}.
\end{eqnarray} 
Here we omit the Breit-Wigner function for simplicity.

\noindent
\textit{Step 3}: For each $\tilde{N}$ random momentum $P^*_j$, generate random variables $(\Theta^*_j, \Phi^*_j)$ whose ensemble is uniformly distributed on the unit sphere.
By using these random variables, we obtain an ensemble of $\rho$ meson with momentum $\bm{P}^*_j = (P^*_{xj}, P^*_{yj}, P^*_{zj})= (P^*_j \sin\Theta^*_j \cos \Phi^*_j, P^*_j \sin\Theta^*_j \sin \Phi^*_j, P^*_j \cos \Theta^*_j)$, which obeys the Bose-Einstein distribution in the fluid rest system.

\noindent
\textit{Step 4}: Boost $\bm{P}^*_j$ with respect to the fluid velocity $\bm{v}_k$
\begin{eqnarray}
\bm{P}_j = \bm{P}^*_j+\bm{v}_k \gamma_k \left(E^*_j+\frac{\bm{P}^*_j\cdot \bm{v}_k\gamma_k}{1+\gamma_k} \right),
\end{eqnarray}
where $\gamma_k = 1/\sqrt{1-\bm{v}_k^2}$.

\noindent
\textit{Step 5}: Generate $\tilde{N}$ uniform random variables on the unit sphere  $(\theta^*_j, \phi^*_j)$ and obtain an ensemble of negative pions with momentum $\bm{p}^*_j = (p^*_{xj}, p^*_{yj}, p^*_{zj}) = (p^* \sin\theta^*_j \cos \phi^*_j, p^* \sin\theta^*_j \sin \phi^*_j, p^* \cos \theta^*_j)$, where $p^*$ is given by
\begin{eqnarray}
p^* = \frac{1}{2 m_R}\sqrt{(m_R + m_\pi)^2-m_X^2}\sqrt{(m_R - m_\pi)^2-m_X^2}.
\end{eqnarray}
For the decay process $\rho\rightarrow\pi\pi$, $m_X=m_\pi$.

\noindent
\textit{Step 6}: Boost $\bm{p}^*_j$ with respect to the resonance momentum $\bm{P}_j$
\begin{eqnarray}
\bm{p}_j = \bm{p}^*_j+\bm{P}_j\left[\frac{E_j}{m_{R}}+\frac{\bm{p}^*_j \cdot \bm{P}_j}{m_{R}(m_{R}+E_{R})} \right].
\end{eqnarray}

\noindent
\textit{Step 7}: If $P^\mu_j d\sigma_{\mu k}$ is positive, 
\begin{eqnarray}
N^+_k \rightarrow N^+_k + \frac{P^\mu_j d\sigma_{\mu k}}{E_j^*}.
\end{eqnarray}
If $P^\mu_j d\sigma_{\mu k}$ is negative,
\begin{eqnarray}
N^-_k \rightarrow N^-_k + \frac{\mid P^\mu_j d\sigma_{\mu k} \mid}{E_j^*}.
\end{eqnarray}
Here, $N^+_k$ ($N^-_k$) is to be proportional to the number of $\rho$ mesons which are emitted from (absorbed by) the $k$-th fluid element.

\noindent
\textit{Step 8}: If the rapidity of a negative pion $Y_j$ which is evaluated from $\bm{p}_j$ enters in a rapidity window $Y-\frac{\Delta Y}{2} < Y_j < Y+\frac{\Delta Y}{2}$ and $P^\mu_j d\sigma_{\mu k}$ is positive,
\begin{eqnarray}
\Delta N^+_k(Y) \rightarrow \Delta N^+_k (Y)+\frac{P^\mu_j d\sigma_{\mu k}}{E_j^*}.
\end{eqnarray}
If $Y_j$ also enters the above rapidity window but $P^\mu_j d\sigma_{\mu k}$ is negative,
\begin{eqnarray}
\Delta N^-_k(Y) \rightarrow \Delta N^-_k(Y)+\frac{\mid P^\mu_j d\sigma_{\mu k} \mid}{E_j^*}.
\end{eqnarray}

\noindent
\textit{Step 9}: Repeat steps 7 and 8 for all $\tilde{N}$ random variables.

\noindent
\textit{Step 10}: Obtain the rapidity distribution of decay particles from the $k$-th
fluid element
\begin{eqnarray}
\label{DNKDY}
\frac{dN_k}{dY}(Y) = \frac{N^R_k}{N^+_k + N^-_k}\left[\frac{\Delta N^+_k(Y) - \Delta N^-_k(Y)}{\Delta Y} \right].
\end{eqnarray}
It should be noted that the minus sign in the bracket of Eq.~(\ref{DNKDY}) means the \textit{net} number of emitted particles from the $k$-th fluid element, which is consistent with the Cooper-Frye prescription.
For the normalization in Eq.~(\ref{DNKDY}), we use the gross number $N_k^+ + N_k^-$ since this number is positive definite.

\noindent
\textit{Step 11}: Repeat the above steps from 1 to 10 for all fluid elements obtained in a numerical simulation of the hydrodynamic model.
Summing over the contribution from all fluid elements on the freeze-out hypersurface $\Sigma$, we obtain the rapidity distribution of negative pions which are from $\rho$ decays:
\begin{eqnarray}
\frac{dN_{\rho \rightarrow \pi^- X}}{dY}(Y) = \sum_k \frac{dN_k}{dY}(Y).
\end{eqnarray}

\end{document}